\documentclass[apj,appendixfloats]{emulateapj}
\usepackage{amsmath}
\usepackage{natbib}
\usepackage{graphics}
\usepackage{appendix}
\begin{document}
\title{Optical Phase Curves of Kepler Exoplanets}
\author{Lisa J. Esteves}
\affil{Astronomy \& Astrophysics, University of Toronto, 50 St. George Street, Toronto, Ontario, Canada, M5S 3H4}
\email{esteves@astro.utoronto.ca}
\author{Ernst J. W. De Mooij}
\affil{Astronomy \& Astrophysics, University of Toronto, 50 St. George Street, Toronto, Ontario, Canada, M5S 3H4}
\email{demooij@astro.utoronto.ca}
\author{Ray Jayawardhana}
\affil{Astronomy \& Astrophysics, University of Toronto, 50 St. George Street, Toronto, Ontario, Canada, M5S 3H4}
\email{rayjay@astro.utoronto.ca}
\begin{abstract}
\indent We have conducted a comprehensive search for optical phase variations of all planet candidates with tight orbits ($a/R_{\star}<$10) in fifteen quarters of data from the {\it Kepler} space telescope. After correcting for systematics, we found eight systems that appear to show secondary eclipses as well as phase variations. Of these, five (Kepler-5, Kepler-6, Kepler-8, KOI-64 and KOI-2133) are new and three (TrES-2, HAT-P-7 and KOI-13) have previously published phase curves, albeit with many fewer observations. We model the full phase curve of each planet candidate, including the primary and secondary transits, and derive their albedos, day- and night-side temperatures, ellipsoidal variations and Doppler beaming. We find that KOI-64 and KOI-2133 have night-side temperatures well above their equilibrium values (while KOI-2133 also has an albedo $>$1), so we conclude that they are likely to be self-luminous objects rather than planets. The characteristics of the six other candidates are consistent with their being planets with low geometric albedos ($<$0.3). For TrES-2 and KOI-13, the {\it Kepler} bandpass appears to probe atmospheric layers hotter than the planet's equilibrium temperature. For KOI-13, we detect a never-before-seen third cosine harmonic with an amplitude of $6.7\pm0.3$ ppm and a phase shift of $-1.1\pm0.1$ radians in the phase curve residual, which could be due to its spin-orbit misalignment. We report derived planetary parameters for all six planets, including masses from ellipsoidal variations and Doppler beaming, and compare our results to published values when available. Our results nearly double the number of {\it Kepler} exoplanets with measured phase curve variations, thus providing valuable new constraints on the properties of close-in hot Jupiters.
\end{abstract}
\maketitle
\section{Introduction}\label{sec:intro}
\indent Until recently, most measurements of the day-side emission of hot Jupiters have relied on targeting the secondary eclipses of these planets. Typically, these studies have focused on the thermal emission in the near- and mid-infrared using the {\it Spitzer} Space Telescope (e.g. review by~\citeauthor{Deming2009}~\citeyear{Deming2009}) as well as ground-based telescopes (e.g.~\citeauthor{DeMooij2009}~\citeyear{DeMooij2009};~\citeauthor{Croll2010}~\citeyear{Croll2010}). However, such observations only permit indirect measurements of the albedo and the day-night contrast of exoplanets~\citep[e.g.][]{Cowan2011}. Phase curve measurements with {\it Spitzer}, on the other hand, have provided direct measurements of the day-night contrasts~\citep[e.g.][]{Knutson2007}, thus the temperature difference between the two hemispheres, and have shown that the hottest spot in the planet's atmosphere could be offset from sub-stellar point~\citep[e.g.][]{Knutson2007}. \\
\indent At optical wavelengths, reflected light could account for a significant fraction of a planet's light curve. Moreover, since the planet-to-star contrast is much lower in the optical regime, contributions from ellipsoidal variations and Doppler boosting also become important. Both these effects provide information on the planet-to-star mass ratio. Ellipsoidal variations stem from changes in the star's light due to tides raised by the planet, while Doppler boosting results from the reflex motion ($K_{\star}$) of the star. So far, optical phase curves of only a handful planets have been presented in the literature: CoRoT-1b~\citep{Snellen2009}, HAT-P-7b (e.g.~\citeauthor{Borucki2009}~\citeyear{Borucki2009};~\citeauthor{Welsh2010}~\citeyear{Welsh2010}), KOI-13 (e.g.~\citeauthor{Shporer2011}~\citeyear{Shporer2011};~\citeauthor{Mazeh2012}~\citeyear{Mazeh2012}), TrES-2b (\citeauthor{Kipping2011}~\citeyear{Kipping2011};~\citeauthor{Barclay2012}~\citeyear{Barclay2012}), and Kepler 41~\citep{Quintana2013}. \\
\indent The {\it Kepler} space telescope monitors over 150,000 stars, and so far the {\it Kepler} team has publicly released fifteen quarters of data, acquired over three years of continuous observations. The majority of stars only have long-cadence (LC) measurements, with a sampling rate of 29.425 minutes, while a small fraction also have short-cadence (SC) observations, with a sampling rate of 58.85 seconds~\citep{Borucki2011}. \\
\indent Here we present the results of our analysis of the first fifteen quarters of {\it Kepler} LC and SC data for eight objects (Kepler-5b, Kepler-6b, Kepler-8b, KOI-13, KOI-64, KOI-2133, TrES-2b, HAT-P-7b) that exhibit phase variations.  In Section~\ref{sec:data} we present the dataset and our analysis method, while in Section~\ref{sec:analysis} we present our model to fit the data. The results are presented and discussed in Section~\ref{sec:res}, and finally we provide the conclusions in Section~\ref{sec:concl}. \\
\section{Data Reduction}\label{sec:data}
\indent After correcting for systematics (see Sec.~\ref{sec:sys} below), we visually inspected the phase curves of all publicly released Kepler planetary candidates and confirmed planets that have a semi-major axis to stellar radius ($a/R_{\star}$) ratio of less than 10. Of these, we found 8 systems (Kepler-5, Kepler-6, Kepler-8, KOI-13, KOI-64, KOI-2133, TrES-2 and HAT-P-7) that, after the removal of systematics, exhibited an apparent phase curve signal. \\
\subsection{Removal of Systematics}\label{sec:sys}
\indent In our analysis, we used both the {\it Kepler} LC and SC simple aperture photometry (SAP) data available (see Table~\ref{tab:data}). Instrumental signals were removed by performing a linear least squares fit\footnote{Using custom IDL procedures.} of the first eight cotrending basis vectors (CBVs)~\citep{KeplerHandbook} to the time-series of each quarter individually. Before cotrending, we removed any bad data points flagged by {\it Kepler} in the SAP or CBV files and to prevent contamination we only fit the CBVs to the out-of-transit time-series. The fitted basis vectors were then divided out of the quarter, in order to preserve the amplitude of the physical signals of interest. Since CBVs are only provided for the LC data, we interpolated onto the SC time-stamps using cubic splines. \\
\begin{table}[h]
\centering
\caption{{\it Kepler} Quarters of Data Used in Analysis}
\begin{tabular}{lcc}
\hline
System & SC Quarter & LC Quarters \\
\hline
Kepler-5 & 2, 3, 4, 5, 6, 7, 8, 9 & 0, 1, 13, 14 \\
		 & 10, 11, 12 & --- \\
Kepler-6 & 2, 3, 4, 5, 6, 9, 10 & 0, 1, 8, 14 \\
		 & 11, 12, 13 & \\
Kepler-8 & 2, 3, 4, 5, 6, 7, 9, 10 & 0, 1, 8, 14 \\
		 & 11, 12, 13 & \\
KOI-64 & 3, 4, 5, 6, 7, 8, 9, 10 & 0, 1, 2, 14 \\
		& 11, 12, 13 & \\
KOI-2133 & ---  & 0, 1, 2, 3, 4, 5, 6, 7, 8 \\
		&  & 9, 10, 11, 12, 13, 14 \\
TrES-2 & 0, 1, 2, 3, 5, 6, 7, 9 & --- \\
	 & 10, 11, 13, 14 & \\
HAT-P-7 & 0, 1, 2, 3, 4, 5, 6, 7 & --- \\
	 & 8, 9, 10, 11, 12, 13, 14 & \\
KOI-13 & 2, 3, 7, 8, 9, 10, 11 & 0, 1, 4, 5, 6 \\
		 & 12, 13, 14 & \\
\hline
\end{tabular}
\label{tab:data}
\end{table}
\indent In order to remove quarter-to-quarter discontinuities we normalized each quarter to its out-of-transit median. After cotrending and combining all quarters we removed outliers by calculating a running median and standard deviation of 21 measurements around each point and rejecting measurements that differed by more than 3$\sigma$. We also calculated the out-of-transit median for each half of the planet's orbit, where an orbit is defined as the time between two consecutive transits, and removed any orbits whose median deviated by more than 2$\sigma$. This was done to remove sections of the light curve where the CBV fit poorly without introducing a phase curve sampling bias. The raw, cotrended and cotrended/out-of-transit/outlier-filtered light curve, of each system, can be found in Figs.~\ref{fig:l1}-\ref{fig:l4}. \\
\subsection{Companion Stars}\label{sec:companions}
\indent {\it Kepler's} large pixel size, with a width of 3.98", allows for the possibility of dilution from a background or foreground star or a nearby stellar companion. In the literature, we find that several of our 8 systems have 1 or 2 companion stars within 4" of the planetary host star (see Table~\ref{tab:var}). However, the only system that is significantly diluted by its companion is KOI-13, which we have corrected for as the contamination would greatly affect the derived planetary parameters. Note that each of these systems could also have closer companions that could not be detected by previous studies and that they could significantly dilute our results. \\
\begin{table}[h]
\caption{Detected Stellar Companions around Planet Host Stars}
\centering
\begin{tabular}{lcccc}
\hline
Host Star & Host Star & Comp. & Comp. Est. & Comp. Est. \\
 	& Kp Mag & Dist (") & Kep Mag & Flux \% \\
\hline
Kepler-5 & 13.369 \textsuperscript{b} & 0.9 \textsuperscript{b} & 18.7 & $<$1\% \\
				   & & 3.39 \textsuperscript{b} & 19.8 & $<$1\% \\
Kepler-6 & 13.303 \textsuperscript{b} & --- \textsuperscript{b} & --- & --- \\
Kepler-8 & 13.563 \textsuperscript{b} & 3.04 \textsuperscript{b} & 22.1 & $<$1\% \\
				   & & 3.74 \textsuperscript{b} & 20.5 & $<$1\% \\
KOI-64 & 13.143 \textsuperscript{b} & --- \textsuperscript{b} & --- & --- \\
KOI-2133 & 12.495 \textsuperscript{c} & --- \textsuperscript{a} & --- & --- \\
TrES-2 & 11.338 \textsuperscript{c} & 0.9 \textsuperscript{d} & --- & $<$1\% \\
HAT-P-7 & 10.463 \textsuperscript{c} & --- \textsuperscript{e} & --- & --- \\
KOI-13 & 9.958 \textsuperscript{b} & 1.12 \textsuperscript{b} & 10.5 & 38\% \\
\hline
\end{tabular}
\begin{tabular}{p{9cm}}
\textsuperscript{a} No data is available. \\
\textsuperscript{b} From~\citet{Adams2012}. \\
\textsuperscript{c} From~\citet{Batalha2013}. \\
\textsuperscript{d} From~\citet{Daemgen2009}. \\
\textsuperscript{e} From~\citet{Narita2010}. \\
\end{tabular}
\label{tab:var}
\end{table}
\subsection{Stellar Variability}\label{sec:var}
\indent The periodogram~\citep{Zechmeister2009} of KOI-64 revealed a strong periodic signal, sharply peaked at a period of 2.224 days, with variations in phase and amplitude between oscillations. We modeled the variability of 2.224-day segments using a linear polynomial and a sine wave with a 2.224-day period, while allowing for small shifts in phase between segments. To minimize discontinuities between periods we simultaneously fit half a period on either side of each segment, then stitched the segments together by interpolating cubic splines over the first and last 10\%. The light curve before and after variability removal can be found in the {\it lower} plot of Fig.~\ref{fig:l2}. \\
\indent Periodograms of the other systems showed that, close to the planet's period or aliases of the period, the stellar variability had an amplitude much lower than the phase curve signal. \\ 
\section{Analysis}\label{sec:analysis}
\indent We modeled the transit and phase curve separately and in two stages in order to remove the phase curve baseline from the transit light curve.
\subsection{Transit Modeling}\label{sec:trans}
\indent To model the transit we used a~\citet{MandelAgol} transit model for a quadratically limb darkened source, over an orbital phase of -0.1 to 0.1, which we fit to our data using a Markov Chain Monte Carlo simulation. The simulation simultaneously fit for the impact parameter of the transit ($b$), the semi-major axis of the planet's orbit to star radius ($a/R_{\star}$), the planet to star radius ($r/R_{\star}$) and the linear and quadratic limb-darkening coefficients ($\gamma_1$ and $\gamma_2$). Five sequences of 100,000 steps were generated and the first 30,000 points were trimmed to avoid any contamination from the initial conditions. The chains were then combined after checking that they were well mixed~\citep{Gelman1992}. \\
\indent The transit curve of KOI-13 is asymmetric as a result of the planet's motion across a stellar surface temperature gradient during transit~\citep{Szabo2011}. To obtain a symmetric curve we averaged the transit in 30-second bins, reflected the curve onto itself and took the mean of each bin. Fitting this curve provided a good first order approximation of the transit depth and shape. \\
\subsection{Phase Curve Modeling}\label{sec:phasm}
\indent We modeled the normalized, out-of-transit phase curve as a sum of four contributions: i) $F_{\mathrm{p}}$, the planet's phase function; ii) $F_{\mathrm{ecl}}$, the secondary eclipse, when the light from the planet is blocked as it passes behind its host star; iii) $F_{\mathrm{d}}$, the Doppler boost caused by the host star's changing radial velocity; iv) $F_{\mathrm{e}}$, the ellipsoidal variations resulting from tides on the star raised by the planet. Each of these components is phase ($\phi$) dependent with $\phi$ running from 0 to 1 and mid-transit occurring at $\phi$=0. The change in brightness of the planet-star system as a function of phase can then be described by
\begin{eqnarray}
\frac{\Delta F}{F} = f_0 + F_{\mathrm{ecl}}(\phi) + F_{\mathrm{p}}(\phi) + F_{\mathrm{d}}(\phi) + F_{\mathrm{e}}(\phi)
\end{eqnarray}
where $f_0$ is an arbitrary zero-point in flux. The details of phase curve model fit are the same as described in Sec.~\ref{sec:trans}. \\
\subsection{Secondary Eclipse}\label{sec:eclip}
\indent Since each of these systems appear to have a secondary eclipse centered on $\phi$=0.5, we assume that the orbits have zero eccentricity and model the secondary eclipse using the formalism from~\citet{MandelAgol} for a uniform source.
\begin{table*}[ht]
\centering
\caption{Limb Darkening, Gravity Darkening and Higher-Order Ellipsoidal Coefficients}
\begin{tabular}{lcccccccc}
\hline
& Kepler-5 & Kepler-6 & Kepler-8 & KOI-64 & KOI-2133 & TrES-2 & HAT-P-7 & KOI-13 \\
\hline
$u$ & $0.290$ & $0.398$ & $0.298$ & $0.474$ & $0.549$ & $0.354$ & $0.282$ & $0.624$ \\
$y$ & $0.545$ & $0.628$ & $0.549$ & $0.650$ & $0.733$ & $0.580$ & $0.551$ & $0.476$ \\
$f_1$ & $0.0154$ & $0.0173$ & $0.0139$ & $0.0288$ & $0.0403$ & $0.0142$ & $0.0214$ & $0.0460$ \\
$f_2$ & $0.0259$ & $0.0288$ & $0.0242$ & $0.0622$ & $0.0672$ & $0.0247$ & $0.0378$ & $0.0779$ \\
\hline
\end{tabular}
\label{tab:dark}
\end{table*}
\subsection{Phase Function}\label{sec:phasf}
\indent We model the variation in planetary light as a Lambert sphere~\citep{Russell1916} described by
\begin{eqnarray}
F_{\mathrm{p}} = A_{\mathrm{p}} \frac{\sin z + (\pi-z)\cos z}{\pi}
\end{eqnarray}
where $A_{\mathrm{p}}$ is the amplitude of the phase function and $z$ is related to $\phi$ and the orbital inclination ($i$) through
\begin{eqnarray}
\cos (z) = - \sin (i) \cos (2\pi\phi)
\end{eqnarray}
\subsection{Doppler Boosting}\label{sec:dopp}
\indent Doppler boosting is a combination of a bolometric and a bandpass dependent effect. The bolometric effect is the result of non-relativistic Doppler boosting of the stellar light in the direction of the star's radial velocity. The observed periodic brightness change is proportional to the star's radial velocity, which is a function of the planet's distance and mass~\citep{Barclay2012}. The bandpass dependent effect is a periodic red/blue shift of the star's spectrum, which results in a periodic measured brightness change as parts of the star's spectrum move in and out of the observed bandpass~\citep{Barclay2012}. The amplitude of the Doppler boosting is modeled by
\begin{eqnarray}
F_{\mathrm{d}} = A_{\mathrm{d}} \sin (2\pi\phi)
\end{eqnarray}
where $A_{\mathrm{d}}$ is the amplitude of the Doppler boost. Given that the radial velocities are much lower than the speed of light and that the planet has zero eccentricity, $A_{\mathrm{d}}$ can be parameterized by
\begin{eqnarray}
A_{\mathrm{d}} = \alpha_{\mathrm{d}} \frac{K_{\star}}{c}
\end{eqnarray}
Here, $c$ is the speed of light, $\alpha_{\mathrm{d}}$ is the photon-weighted bandpass-integrated beaming factor and $K_{\star}$ is the radial velocity semi-amplitude given by
\begin{eqnarray}
K_{\star} = \left( \frac{2 \pi G}{P} \right)^{1/3} \frac{M_{\mathrm{p}} \sin i}{M_{\star}^{2/3}}
\end{eqnarray}
where $G$ is the universal gravitational constant, $P$ is the orbital period of the planet and we have assumed $M_{\mathrm{p}}<<M_{\star}$. Similar to~\citet{Barclay2012}, we calculated $\alpha_{\mathrm{d}}$ in the manner described by~\citet{Bloemen2011} and~\citet{Loeb2003}.
\begin{eqnarray}
\alpha_{\mathrm{b}} = \frac{ \int T_{\mathrm{K}} \left( 5+\frac{\mathrm{d}\ln F_{\lambda,\star} } {\mathrm{d}\ln \lambda} \right) \lambda F_{\lambda,\star} \mathrm{d} \lambda} {\int T_{\mathrm{K}} \lambda F_{\lambda,\star} \mathrm{d} \lambda }
\end{eqnarray}
$T_{\mathrm{K}}$ is the Kepler transmission function, $\lambda$ is the wavelength and $F_{\lambda,\star}$ is the stellar flux computed using the NEXTGEN model spectra~\citep{Nextgen1999}. \\
\indent We opted to fit Kepler-5, Kepler-6, and KOI-2133 without Doppler boosting as they exhibit a poorly constrained, negative Doppler signal.
\subsection{Ellipsoidal Variations}\label{sec:elli}
\indent Ellipsoidal variations are periodic changes in observed stellar flux caused by fluctuations of the star's visible surface area as the stellar tide, created by the planet, rotates in and out of view of the observer~\citep{Mislis2012}. If there is no tidal lag, the star's visible surface area and ellipsoidal variations are at maximum when the direction of the tidal bulge is perpendicular to the observer's line of sight and at minimum during the transit and secondary eclipse.\\
\indent The ellipsoidal light curve is described, by Eqs. 1-3 of~\citet{Morris1985}, as a linear combination of the first three cosine harmonics of the planet's period. These equations can be re-expressed as
\begin{eqnarray}
F_{\mathrm{e}} &=& - A_{\mathrm{e}} \left[ \cos(2\pi\cdot2\phi) + f_1\cos(2\pi\phi) + f_2\cos(2\pi\cdot3\phi) \right]  \nonumber \\
&&
\end{eqnarray}
$A_{\mathrm{e}}$ is the amplitude of the dominant cosine harmonic and $f_1$ and $f_2$ are fractional constants defined by 
\begin{eqnarray}
f_1 &=& 3  \alpha_1 \left( \frac{a}{R_{\star}} \right)^{-1} \frac{5 \sin^2 i - 4}{\sin i} \\
f_2 &=& 5  \alpha_1 \left( \frac{a}{R_{\star}} \right)^{-1} \sin i
\end{eqnarray}
$A_{\mathrm{e}}$ is parameterized as
\begin{eqnarray}
A_{\mathrm{e}} =  \alpha_2 \frac{M_{\mathrm{p}}}{M_{\star}} \left( \frac{a}{R_{\star}} \right)^{-3} \sin^2 i
\end{eqnarray}
where $M_{\star}$ is the mass of the star and $M_{\mathrm{p}}$ is the mass of the planet; the only free parameter in our fit of the ellipsoidal variations. The constants $\alpha_1$ and $\alpha_2$ are defined as
\begin{eqnarray}
\alpha_1 &=& \frac{25 u}{24 (15+u)} \left( \frac{y+2}{y+1} \right) \\
\alpha_2 &=& \frac{3(15+u)}{20(3-u)} (y+1)
\end{eqnarray}
where $u$ and $y$ are the linear limb darkening and gravity darkening parameters, respectively. Similar to~\citet{Barclay2012} we trilinearly interpolate for $u$ and $y$ calculated by~\citet{Claret2011} from the grids in effective temperature, surface gravity and metallicity using the {\it Kepler} filter, a microturbulent velocity of 2 km s$^{-1}$ and ATLAS model spectra (See Table~\ref{tab:dark}).
\section{Results and Discussion}\label{sec:res}
\begin{table*}[ht]
\centering
\caption{Stellar and Planetary Parameters}
\setlength{\tabcolsep}{0.1cm}
\begin{tabular}{lcccc}
\hline
Parameter & Kepler-5 & Kepler-6 & Kepler-8 & KOI-64 \\
\hline
\hline \\ [-2.0ex]
Period (Days) \textsuperscript{a} & 
$3.5484657\pm0.0000007$   & 
$3.2346995\pm0.0000004$   & 
$3.522297\pm0.0000007$   & 
$1.9510914\pm0.000004$  \\
$T_0$ (BJD-2454900) \textsuperscript{a} & 
$55.90078\pm0.00007$   & 
$54.48580\pm0.00004$   & 
$54.11860\pm0.00062$   & 
$90.54077\pm0.00052$  \\
$T_{\star}$ (K) & 
$6297\pm60$ \textsuperscript{b}  & 
$5647\pm44$ \textsuperscript{c}  & 
$6213\pm150$ \textsuperscript{d}  & 
$5128\pm200$ \textsuperscript{a} \\
$\log g$ (cgs) & 
$3.96\pm0.10$ \textsuperscript{b}  & 
$4.236\pm0.011$ \textsuperscript{c}  & 
$4.28\pm0.10$ \textsuperscript{d}  & 
$3.94\pm0.5$ \textsuperscript{a} \\
$\mathrm{[Fe/H]}$ & 
$0.04\pm0.06$ \textsuperscript{b}  & 
$0.34\pm0.04$ \textsuperscript{c}  & 
$-0.055\pm0.03$ \textsuperscript{d}  & 
$-0.341\pm0.5$ \textsuperscript{a} \\
$R_{\star}/R_{\sun}$ & 
$1.793^{+0.043}_{-0.062}$ \textsuperscript{b}  & 
$1.391^{+0.017}_{-0.034}$ \textsuperscript{c}  & 
$1.486^{+0.053}_{-0.062}$ \textsuperscript{d}  & 
$1.938$ \textsuperscript{a} \\
$M_{\star}/M_{\sun}$ & 
$1.374^{+0.040}_{-0.059}$ \textsuperscript{b}  & 
$1.209^{+0.044}_{-0.038}$ \textsuperscript{c}  & 
$1.213^{+0.067}_{-0.063}$ \textsuperscript{d}  & 
$1.19$ \textsuperscript{a} \\
\hline
\multicolumn{5}{l}{Transit Fit} \\
\hline \\ [-2.0ex]
$R_{\mathrm{p}}/R_{\star}$ & 
$0.078845^{+0.000047}_{-0.000056}$   & 
$0.092853^{+0.000037}_{-0.000045}$   & 
$0.094337^{+0.000099}_{-0.000085}$   & 
$0.04038^{+0.00040}_{-0.00051}$  \\
$a/R_{\star}$ & 
$6.365^{+0.019}_{-0.014}$   & 
$7.5606^{+0.0034}_{-0.0031}$   & 
$6.820^{+0.017}_{-0.018}$   & 
$3.972^{+0.070}_{-0.078}$  \\
$b$ & 
$0.188^{+0.011}_{-0.017}$   & 
$0.032^{+0.011}_{-0.013}$   & 
$0.7212\pm0.0021$   & 
$0.9324^{+0.0049}_{-0.0039}$  \\
$i$ (degrees) & 
$88.31^{+0.16}_{-0.10}$   & 
$89.759^{+0.099}_{-0.082}$   & 
$83.929^{+0.033}_{-0.034}$   & 
$76.42^{+0.29}_{-0.35}$  \\
$\gamma_1$ & 
$0.3494^{+0.0039}_{-0.0032}$   & 
$0.4691^{+0.0034}_{-0.0073}$   & 
$0.305^{+0.023}_{-0.014}$   & 
$0.466^{+0.045}_{-0.035}$  \\
$\gamma_2$ & 
$0.1711^{+0.0044}_{-0.0052}$   & 
$0.1762^{+0.0176}_{-0.0059}$   & 
$0.252^{+0.012}_{-0.032}$   & 
$0.306^{+0.041}_{-0.039}$  \\
\hline
\multicolumn{5}{l}{Phasecurve Fit}\\
\hline \\ [-2.0ex]
$F_{\mathrm{ecl}}$ (ppm) & 
$18.8\pm3.7$   & 
$8.9\pm3.8$   & 
$26.2\pm5.6$   & 
$61.4\pm3.8$  \\
$F_{\mathrm{n}}$ (ppm) & 
$2\pm4$   & 
$-4\pm4$   & 
$0.8\pm6$   & 
$49\pm4$  \\
$A_{\mathrm{p}}$ (ppm) & 
$16.5\pm2.0$   & 
$12.4\pm2.0$   & 
$25.3^{+2.7}_{-2.6}$   & 
$12.5^{+1.8}_{-1.9}$  \\
$A_{\mathrm{d}}$ (ppm) & 
$$ --- $$   & 
$$ --- $$   & 
$2.5\pm1.2$   & 
$3.05\pm0.80$  \\
$A_{\mathrm{e}}$ (ppm) & 
$4.7^{+1.0}_{-1.1}$   & 
$2.7\pm1.0$   & 
$4.0\pm1.4$   & 
$15.20\pm0.93$  \\
\hline
\multicolumn{5}{l}{Derived Parameters}\\
\hline \\ [-2.0ex]
$R_{\mathrm{p}}$ $({\mathrm{R_J}})$ & 
$1.406^{+0.034}_{-0.049}$   & 
$1.285^{+0.016}_{-0.031}$   & 
$1.395^{+0.050}_{-0.058}$   & 
$0.779\pm0.041$  \\
$a$ (Au) & 
$0.0531^{+0.0013}_{-0.0018}$   & 
$0.04889^{+0.00060}_{-0.00120}$   & 
$0.0471^{+0.0017}_{-0.0020}$   & 
$0.0358^{+0.0019}_{-0.0020}$  \\
$M_{\mathrm{p}}$ from $A_{\mathrm{d}}$ $({\mathrm{M_J}})$ & 
$$ --- $$   & 
$$ --- $$   & 
$1.85^{+0.90}_{-0.88}$   & 
$1.52\pm0.41$  \\
$M_{\mathrm{p}}$ from $A_{\mathrm{e}}$ $({\mathrm{M_J}})$ & 
$1.34^{+0.30}_{-0.31}$   & 
$1.02\pm0.40$   & 
$1.23\pm0.43$   & 
$0.829^{+0.097}_{-0.099}$  \\
Weighted $M_{\mathrm{p}}$ $({\mathrm{M_J}})$ & 
$$ --- $$   & 
$$ --- $$   & 
$1.35\pm0.39$   & 
$0.867\pm0.095$  \\
$A_{\mathrm{g,ecl}}$ & 
$0.122\pm0.024$   & 
$0.059\pm0.025$   & 
$0.137\pm0.029$   & 
$0.594^{+0.042}_{-0.044}$  \\
$T_{\mathrm{eq,max}}$ (K) & 
$2260$   & 
$1860$   & 
$2150$   & 
$2320$  \\
$T_{\mathrm{eq,hom}}$ (K) & 
$1760$   & 
$1450$   & 
$1680$   & 
$1820$  \\
$T_{\mathrm{B,day}}$ (K) & 
$2400^{+50}_{-60}$   & 
$2000^{+80}_{-100}$   & 
$2370^{+50}_{-70}$   & 
$2940^{+20}_{-30}$  \\
[1.0ex]
$T_{\mathrm{B,night}}$ (K) & 
$^{<2100(1\sigma)}_{<2300(3\sigma)}$   & 
$^{<1600(1\sigma)}_{<2000(3\sigma)}$   & 
$^{<2100(1\sigma)}_{<2300(3\sigma)}$   & 
$2850\pm30$  \\
\hline
\end{tabular}
\begin{tabular}{p{15cm}}
\textsuperscript{a} From \citet{Batalha2013}.\\
\textsuperscript{b} From \citet{Koch2010}.\\
\textsuperscript{c} From \citet{Dunham2010}.\\
\textsuperscript{d} From \citet{Jenkins2010}.\\
Note: A stellar mass uncertainty of $\pm$0.1M$_{\odot}$ and a stellar radius uncertainity of $\pm$0.1R$_{\odot}$ was assumed when not given in the literature.
\label{tab:res1}
\end{tabular}
\end{table*}
\begin{table*}[ht]
\centering
\caption{Stellar and Planetary Parameters}
\setlength{\tabcolsep}{0.1cm}
\begin{tabular}{lccccc}
\hline
Parameter & KOI-2133 & TrES-2 & HAT-P-7 & \multicolumn{2}{c}{KOI-13} \\
\hline
\hline \\ [-2.0ex]
Period (Days) \textsuperscript{a} & 
$6.2465796\pm0.000082$   & 
$2.4706132\pm0.0000001$   & 
$2.2047355\pm0.0000001$  & \multicolumn{2}{c}{
$1.7635877\pm0.000001$  } \\
$T_0$ (BJD-2454900) \textsuperscript{a} & 
$69.39661\pm0.0048$   & 
$55.76257\pm0.00001$   & 
$54.35780\pm0.00002$  & \multicolumn{2}{c}{
$53.56513\pm0.00001$  } \\
$T_{\star}$ (K) & 
$4712\pm200$ \textsuperscript{a}  & 
$5850\pm50$ \textsuperscript{e}  & 
$6350\pm80$ \textsuperscript{f} & \multicolumn{2}{c}{
$8511\pm1$ \textsuperscript{g} } \\
$\log g$ (cgs) & 
$2.852\pm0.5$ \textsuperscript{a}  & 
$4.4\pm0.1$ \textsuperscript{e}  & 
$4.07^{+0.04}_{-0.08}$ \textsuperscript{f} & \multicolumn{2}{c}{
$3.9\pm0.1$ \textsuperscript{g} } \\
$\mathrm{[Fe/H]}$ & 
$0.509\pm0.5$ \textsuperscript{a}  & 
$-0.15\pm0.10$ \textsuperscript{e}  & 
$0.26\pm0.08$ \textsuperscript{f} & \multicolumn{2}{c}{
$0.2$ \textsuperscript{g} } \\
$R_{\star}/R_{\sun}$ & 
$7.488$ \textsuperscript{a}  & 
$1.000^{+0.036}_{-0.033}$ \textsuperscript{e}  & 
$1.84^{+0.23}_{-0.11}$ \textsuperscript{f} & \multicolumn{2}{c}{
$2.55$ \textsuperscript{g} } \\
$M_{\star}/M_{\sun}$ & 
$2.25$ \textsuperscript{a}  & 
$0.980\pm0.062$ \textsuperscript{e}  & 
$1.47^{+0.08}_{-0.05}$ \textsuperscript{f} & \multicolumn{2}{c}{
$2.05$ \textsuperscript{g} } \\
\hline
\multicolumn{6}{l}{Transit Fit} \\
\hline \\ [-2.0ex]
$R_{\mathrm{p}}/R_{\star}$ & 
$0.01775^{+0.00042}_{-0.00065}$   & 
$0.125106^{+0.000025}_{-0.000024}$   & 
$0.077490\pm0.000013$  & \multicolumn{2}{c}{
$0.080509^{+0.000033}_{-0.000048}$  } \\
$a/R_{\star}$ & 
$4.51^{+0.12}_{-0.26}$   & 
$7.8957^{+0.0028}_{-0.0027}$   & 
$4.1512^{+0.0025}_{-0.0026}$  & \multicolumn{2}{c}{
$4.3396^{+0.0102}_{-0.0075}$  } \\
$b$ & 
$0.0^{+0.19}_{-0.26}$   & 
$0.84388^{+0.00020}_{-0.00026}$   & 
$0.4973^{+0.0011}_{-0.0010}$  & \multicolumn{2}{c}{
$0.3681^{+0.0041}_{-0.0064}$  } \\
$i$ (degrees) & 
$89.9^{+3.3}_{-2.5}$   & 
$83.8646^{+0.0041}_{-0.0036}$   & 
$83.119\pm0.019$  & \multicolumn{2}{c}{
$85.135^{+0.097}_{-0.063}$  } \\
$\gamma_1$ & 
$0.69\pm0.12$   & 
$0.3529^{+0.0024}_{-0.0021}$   & 
$0.3522^{+0.0012}_{-0.0010}$  & \multicolumn{2}{c}{
$0.3047^{+0.0033}_{-0.0038}$  } \\
$\gamma_2$ & 
$0.05^{+0.25}_{-0.12}$   & 
$0.2635^{+0.0031}_{-0.0030}$   & 
$0.1705^{+0.0010}_{-0.0019}$  & \multicolumn{2}{c}{
$0.2249^{+0.0072}_{-0.0063}$  } \\
\hline
\multicolumn{6}{l}{Phasecurve Fit}\\
\hline \\ [-2.0ex]
$F_{\mathrm{ecl}}$ (ppm) & 
$38.7\pm8.2$   & 
$7.5\pm1.7$   & 
$68.31\pm0.69$   & 
$147.24\pm0.82$   & 
$143.0^{+1.2}_{-1.4}$   \\
$F_{\mathrm{n}}$ (ppm) & 
$30\pm10$   & 
$3\pm2$   & 
$2.6\pm0.8$   & 
$19\pm1$   & 
$17^{+1}_{-2}$   \\
$A_{\mathrm{p}}$ (ppm) & 
$13.1^{+5.8}_{-6.0}$   & 
$4.77^{+0.65}_{-0.63}$   & 
$65.75\pm0.48$   & 
$128.67^{+0.59}_{-0.58}$   & 
$125.96^{+0.82}_{-0.91}$   \\
$A_{\mathrm{d}}$ (ppm) & 
$$ --- $$   & 
$2.40\pm0.30$   & 
$5.80\pm0.19$   & 
$7.14\pm0.24$   & 
$7.23^{+0.25}_{-0.24}$   \\
$A_{\mathrm{e}}$ (ppm) & 
$45.2\pm3.1$   & 
$3.67\pm0.33$   & 
$19.09\pm0.25$   & 
$61.28\pm0.31$   & 
$60.97\pm0.32$   \\
$A_{3\phi}$ (ppm) & 
---   & 
---   & 
---   & 
---   & 
$6.71\pm0.26$   \\
$\theta_{3\phi}$ (rad) & 
---   & 
---   & 
---   & 
---   & 
$-1.119^{+0.096}_{-0.148}$   \\
\hline
\multicolumn{6}{l}{Derived Parameters}\\
\hline \\ [-2.0ex]
$R_{\mathrm{p}}$ $({\mathrm{R_J}})$ & 
$1.322^{+0.036}_{-0.051}$   & 
$1.245^{+0.045}_{-0.041}$   & 
$1.418^{+0.177}_{-0.085}$  & \multicolumn{2}{c}{
$2.042\pm0.080$  } \\
$a$ (Au) & 
$0.1569^{+0.0047}_{-0.0091}$   & 
$0.0367^{+0.0013}_{-0.0012}$   & 
$0.0355^{+0.0044}_{-0.0021}$  & \multicolumn{2}{c}{
$0.0514\pm0.0020$  } \\
$M_{\mathrm{p}}$ from $A_{\mathrm{d}}$ $({\mathrm{M_J}})$ & 
$$ --- $$   & 
$1.28\pm0.17$   & 
$4.25^{+0.21}_{-0.17}$   & 
$8.49\pm0.40$   & 
$8.61^{+0.41}_{-0.40}$   \\
$M_{\mathrm{p}}$ from $A_{\mathrm{e}}$ $({\mathrm{M_J}})$ & 
$5.92^{+0.68}_{-1.12}$   & 
$1.37\pm0.15$   & 
$1.631^{+0.091}_{-0.060}$   & 
$7.45\pm0.37$   & 
$7.41\pm0.37$   \\
Weighted $M_{\mathrm{p}}$ $({\mathrm{M_J}})$ & 
$$ --- $$   & 
$1.33\pm0.11$   & 
$1.985\pm0.070$   & 
$7.93\pm0.27$   & 
$7.95\pm0.27$   \\
$A_{\mathrm{g,ecl}}$ & 
$2.49^{+0.55}_{-0.60}$   & 
$0.0301\pm0.0069$   & 
$0.1960\pm0.0020$   & 
$0.4278^{+0.0031}_{-0.0028}$   & 
$0.4153^{+0.0040}_{-0.0043}$   \\
$T_{\mathrm{eq,max}}$ (K) & 
$2009$   & 
$1880$   & 
$2820$   & 
$3690$   & 
$3690$   \\
$T_{\mathrm{eq,hom}}$ (K) & 
$1570$   & 
$1470$   & 
$2200$   & 
$2890$   & 
$2890$   \\
$T_{\mathrm{B,day}}$ (K) & 
$3300\pm100$   & 
$1910^{+40}_{-50}$   & 
$2846\pm4$   & 
$3724\pm3$   & 
$3706^{+5}_{-6}$   \\
$T_{\mathrm{B,night}}$ (K) & 
$3100\pm200$   & 
$1700^{+80}_{-200}$   & 
$1950^{+60}_{-70}$   & 
$2740\pm20$   & 
$2710^{+30}_{-40}$   \\
\hline
\end{tabular}
\begin{tabular}{p{15cm}}
\textsuperscript{a} From \citet{Batalha2013}.\\
\textsuperscript{e} From \citet{Szabo2011}.\\
\textsuperscript{f} From \citet{Sozzetti2007}.\\
\textsuperscript{g} From \citet{Pal2008}. \\
Note: A stellar mass uncertainty of $\pm$0.1M$_{\odot}$ and a stellar radius uncertainity of $\pm$0.1R$_{\odot}$ was assumed when not given in the literature and for KOI-13, the {\it right} column contains results from a model fit including the $3\phi$ term, while the {\it left} column is without. \\
\end{tabular}
\label{tab:res2}
\end{table*}
\begin{figure*}[ht]
\begin{center}
\scalebox{0.8}{\includegraphics{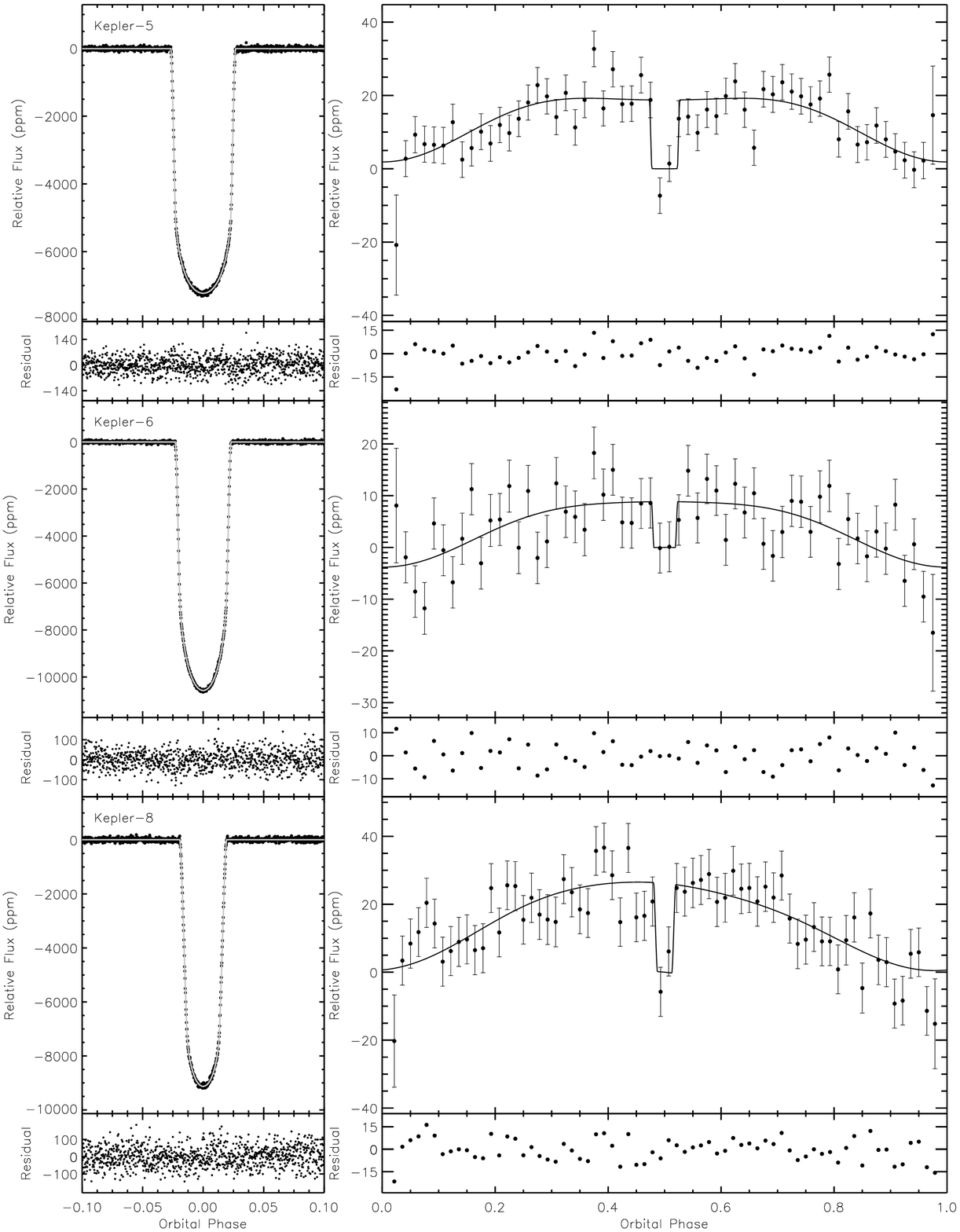}}
\end{center}
\caption{The {\it left} and {\it right} panels contain the binned and phase-folded transit light curves and phase curves, respectively. Over-plotted on each is our best fit model with the residual plotted underneath. For Kepler-5, Kepler-6 and Kepler-8 the transit bin size is 30 seconds while the phase curve bin sizes are 85, 78 and 72 minutes, respectively.}
\label{fig:fig1}
\end{figure*}
\begin{figure*}[ht]
\begin{center}
\scalebox{0.8}{\includegraphics{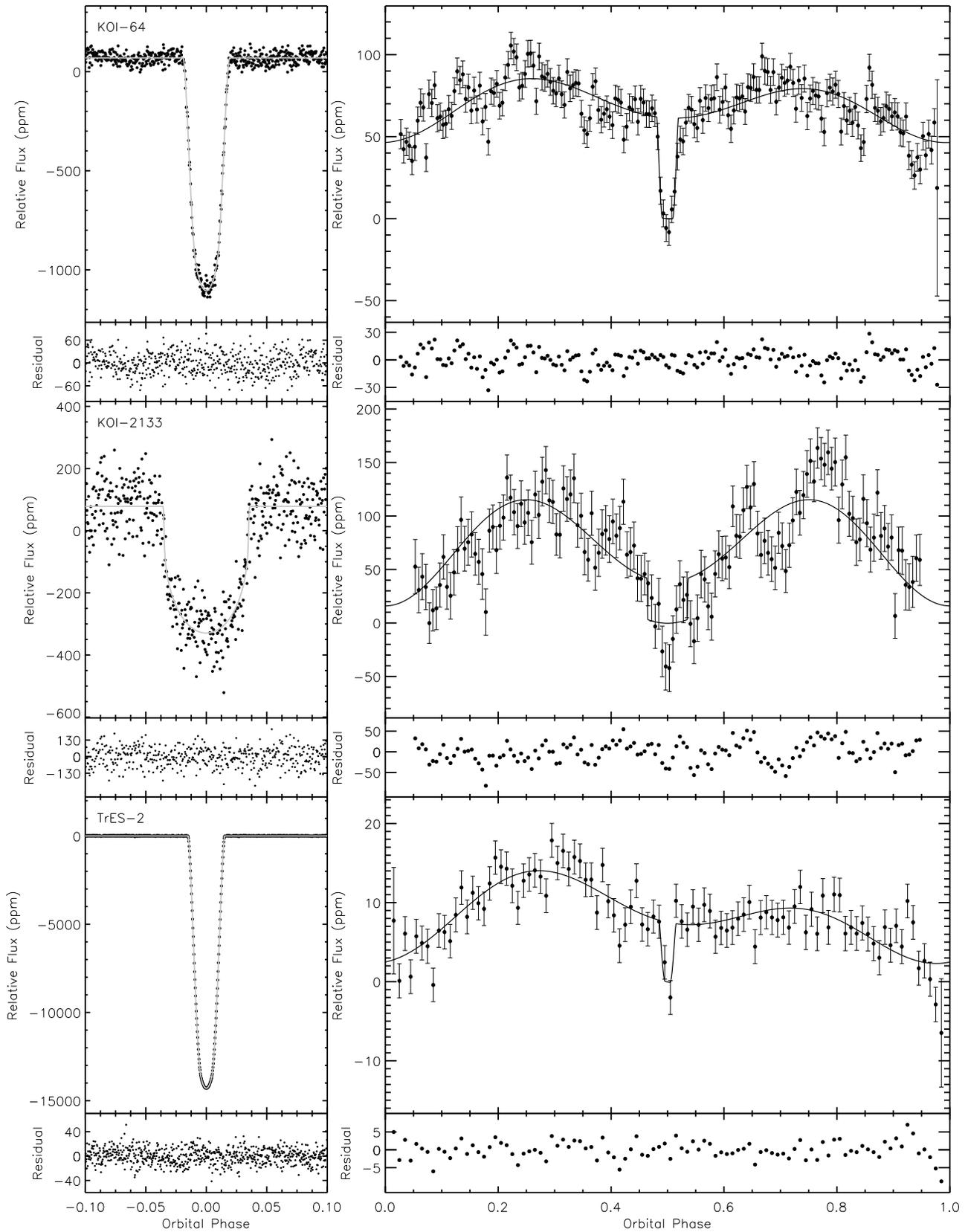}}
\end{center}
\caption{Same as~\ref{fig:fig1}. However, for KOI-64, KOI-2133 and TrES-2 the transit bin sizes are 30, 120 and 30 seconds, respectively, while the phase curve bin sizes are 14, 56 and 30 minutes, respectively.}
\label{fig:fig2}
\end{figure*}
\begin{figure*}[ht]
\begin{center}
\scalebox{0.8}{\includegraphics{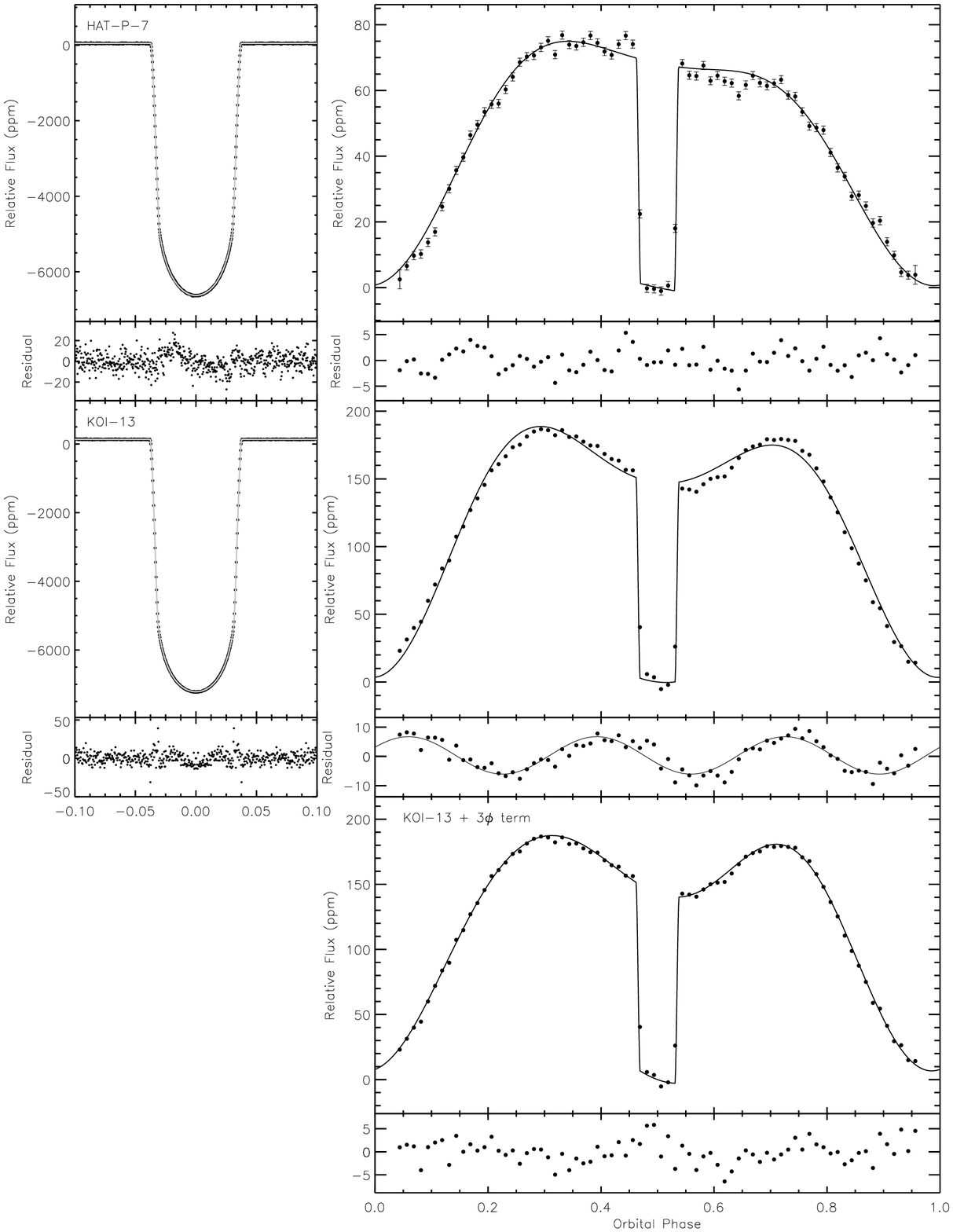}}
\end{center}
\caption{Same as~\ref{fig:fig1}. However, for HAT-P-7 and KOI-13 the transit bin size is 30 seconds while the phase curve bin sizes are 30 and 32 minutes, respectively. In addition, over-plotted on KOI-13's ({\it middle right} panel) residual is the 3$\phi$ signal described in Sec.~\ref{sec:3phi}. The {\it lower right} panel contains the best fit model and residual for a model fit including this additional signal.}
\label{fig:fig3}
\end{figure*}
\indent The relevant stellar, fitted and derived parameters can be found in Tables~\ref{tab:res1}-\ref{tab:res2} and plots of the transit and phase curve fit and residuals, for each system, can be found in Figs.~\ref{fig:fig1}-\ref{fig:fig3}.
\subsection{Derived Masses}\label{sec:mass}
\indent We compared the Kepler-5, Kepler-6, Kepler-8, TrES-2 and HAT-P-7 mass values from radial velocity measurements to the planet masses derived from ellipsoidal variations (see Tables~\ref{tab:res1}-\ref{tab:res2}). We find that TrES-2~\citep{Odonovan2006} and HAT-P-7~\citep{Pal2008} agree with our ellipsoidal mass, while Kepler-6~\citep{Dunham2010} and Kepler-8~\citep{Jenkins2010} are 2$\sigma$ lower and Kepler-5~\citep{Koch2010} is 3$\sigma$ higher than our value. \\
\indent Of these planets, we also derived planet masses from the Doppler boosting signal for Kepler-8, TrES-2 and HAT-P-7 (see Tables~\ref{tab:res1}-\ref{tab:res2}). We find that Kepler-8's Doppler mass is consistent with zero while HAT-P-7's is over 9$\sigma$ higher than its mass from ellipsoidal and radial velocity measurements. TrES-2's is in agreement with both. \\
\indent We also compare our ellipsoidal and Doppler measurements with the previously published phase curves of TrES-2, HAT-P-7 and KOI-13. \\
\indent For TrES-2, our ellipsoidal and Doppler amplitudes agree within 1$\sigma$ to values in~\citet{Barclay2012} and 2$\sigma$ to values in~\citet{Kipping2011}. \\
\indent For HAT-P-7,~\citet{Jackson2012} gives a planet to stellar mass ratio of $(1.10\pm0.06)\cdot 10^{-3}$ and a radial velocity semi-amplitude of $300\pm70$ m s$^{-1}$. Using our formalism this corresponds to $A_{\mathrm{e}}=20\pm1$ ppm and $A_{\mathrm{d}}=3.4\pm0.8$, which are within 1$\sigma$ and 3$\sigma$ of our values, respectively. In addition,~\citet{Mislis2012} find an ellipsoidal and Doppler amplitude of 31 ppm and 8.7 ppm, respectively, while~\citet{Welsh2010} measure $A_{\mathrm{e}}=37.3$. These values are approximately double ours, however this is because~\citet{Mislis2012} and~\citet{Welsh2010} measure peak-to-peak amplitudes, while we measure semi-amplitudes. Another study,~\citet{VanEylen2012}, measured an ellipsoidal amplitude of $59\pm1$, however their model, compared to ours, includes an additional factor of $\pi$. If we take this into account we find that our values agree. \\
\indent For KOI-13,~\citet{Mazeh2012} and~\citet{Shporer2011} find ellipsoidal values of $66.8\pm1.6$ and $30.25\pm0.63$ ppm, respectively, and Doppler values of $8.6\pm1.1$ and $5.28\pm0.44$ ppm, respectively. Note that~\citet{Shporer2011} do not correct for the dilution from KOI-13's companion star and as a result calculate much lower values. From their phase curve analysis,~\citet{Mislis2012KOI13} give a planet mass of $8.3\pm1.25$ M$_{\mathrm{J}}$, which is in agreement with our derived mass. \\
\indent Each of these studies use a different number of observations, systematic removal method and phase curve model. In particular the choice of phase function will influence the derived ellipsoidal mass. As described in~\citet{Mislis2012}, there is a degeneracy between the choice of phase function and amplitude of the ellipsoidal variations. Choosing a wider phase function, such as a geometrical sphere, will result in a lower ellipsoidal amplitude. \\
\subsection{3$\phi$ Signal}\label{sec:3phi}
\indent It is very clear that there is a 3$\phi$ signal present in the phase curve residual of KOI-13 (see Fig.~\ref{fig:fig3}, {\it middle} panel). We have re-modeled KOI-13's phase curve to include the 3$\phi$ cosine signal (see Fig.~\ref{fig:fig3}, {\it lower} panel) and found a significant amplitude ($A_{3\phi}$=$6.7\pm0.3$ ppm) and phase shift ($\theta_{3\phi}$=$-1.1\pm0.1$ radians). Note that this also slightly changed the fitted phase curve parameters (see Table~\ref{tab:res2}). \\
\indent The host star of KOI-13 is a rapid rotator ($v \sin i=65$ km s$^{-1}$), and therefore has significant gravity darkening at the equator compared to the star's poles. This is clearly seen in the asymmetry in the transit caused by a spin-orbit misalignment (\citeauthor{Szabo2011}~\citeyear{Szabo2011};~\citeauthor{Barnes2011}~\citeyear{Barnes2011}). This signal, at three times the orbital frequency, could be due to the tidal bulge caused by the planet, moving across areas with different surface brightnesses. \\
\subsection{Secondary Eclipse and Planetary Phase-function}\label{sec:alb}
\indent For all the systems we detect a significant secondary eclipse and phase function and for KOI-13, KOI-64, KOI-2133 and HAT-P-7 we also detect a significant night-side flux ($F_{\mathrm{n}}$) defined as
\begin{eqnarray}
F_{\mathrm{n}}=F_{\mathrm{ecl}}-A_{\mathrm{p}}
\end{eqnarray}
where $F_{\mathrm{ecl}}$ is the depth of the eclipse and $A_{\mathrm{p}}$ is the amplitude of the phase function (see Tables~\ref{tab:res1}-\ref{tab:res2}). \\
\indent All systems, except KOI-2133 and Kepler-8, have a published secondary eclipse detection of greater than 1$\sigma$. Of these, KOI-13, TrES-2 and HAT-P-7 also have published phase functions and therefore night-side flux measurements. \\
\indent For TrES-2, our measurements agree with the secondary eclipse and phase function values presented in~\citet{Barclay2012} and~\citet{Kipping2011}. \\
\indent For HAT-P-7, the secondary eclipse and phase function values in the literature differ significantly from each other. Our values agree with~\citet{Morris2013} and~\citet{Coughlin2012} and are within 4$\sigma$ of the values presented in~\citet{Jackson2012}, and~\citet{VanEylen2012}. In addition,~\citet{Borucki2009}, who analyze 10 days of data, measure $F_{\mathrm{ecl}}=130\pm11$ ppm and $A_{\mathrm{p}}=122$ ppm, while~\citet{Welsh2010} use 34 days of data and find $F_{\mathrm{ecl}}=85.8$ ppm and $A_{\mathrm{p}}=63.7$ ppm. The large discrepancy between these two studies and our analysis, which includes over 1000 days of data, is most likely due to the number of observations used. \\
\indent For KOI-13, the secondary eclipse values from~\citet{Szabo2011} and~\citet{Coughlin2012} are within 2$\sigma$ of our value. While \citet{Mazeh2012} measure $F_{\mathrm{ecl}}=163.8\pm3.8$ ppm, 4$\sigma$ higher than our value, and a phase function semi-amplitude of $72\pm1.5$ ppm, which, if converted to a peak-to-peak amplitude, is a 8$\sigma$ higher than ours. In addition,~\citet{Shporer2011} measure a phase function semi-amplitude of $39.78\pm0.52$, approximately half our semi-amplitude, due to not removing the dilution from KOI-13's companion. \\
\indent The published eclipse depths of Kepler-5~\citep{Desert2011} and KOI-64~\citep{Coughlin2012} agree with our values while~\citet{Desert2011}, who also examined Kepler-6, using Q0-5 of {\it Kepler} pre-search data conditioned (PDC) data, found an eclipse depth of $22\pm7$, more than double ours. However, our analysis of Kepler-6 includes an additional eight quarters of data and uses cotrended SAP data, which exhibits fewer residual systematics when compared to PDC data~\citep{Still2012}.
\subsection{Planetary Temperatures and Albedos}\label{sec:alb}
\begin{table*}[ht]
\centering
\caption{Self-Consistent Albedos and Temperatures}
\setlength{\tabcolsep}{0.1cm}
\begin{tabular}{lccccccc}
\hline
Parameter & Kepler-5 & Kepler-6 & Kepler-8 & KOI-64 & TrES-2 & HAT-P-7 & KOI-13 \\
\hline \\ [-2.0ex]
$A_{\mathrm{g,max}}$ & 
$0.065\pm0.031$ & 
$0.038\pm0.028$ & 
$0.098\pm0.035$ & 
$0.59337\pm0.037$ & 
$0.0041^{+0.0076}_{-0.0077}$ & 
$0.0299\pm0.0039$ & 
$0.092^{+0.027}_{-0.036}$\\
$A_{\mathrm{g,hom}}$ & 
$0.119\pm0.025$ & 
$0.058\pm0.025$ & 
$0.134\pm0.030$ & 
$0.59358\pm0.037$ & 
$0.0287^{+0.0069}_{-0.0070}$ & 
$0.1849\pm0.0021$ & 
$0.4031^{+0.0039}_{-0.0046}$\\
Max $A_{\mathrm{g}}$ & 
--- & 
--- & 
--- & 
--- & 
0 & 
$0.261^{+0.059}_{-0.049}$ & 
$0.148^{+0.027}_{-0.023}$\\
\hline
\multicolumn{8}{l}{Derived Temperatures}\\
\hline \\ [-2.0ex]
$T_{\mathrm{B,max}}$ (K) & 
$2198^{+28}_{-29}$ & 
$1829\pm20$ & 
$2066^{+31}_{-32}$ & 
$1340^{+150}_{-220}$ & 
$1878.3\pm5.4$ & 
$2784.1^{+4.3}_{-4.2}$ & 
$3558^{+54}_{-42}$\\
$T_{\mathrm{B,hom}}$ (K) & 
$1681\pm19$ & 
$1420\pm15$ & 
$1590^{+22}_{-23}$ & 
$1050^{+110}_{-170}$ & 
$1456.0\pm4.0$ & 
$2032.0\pm2.2$ & 
$2290.9^{+9.9}_{-8.6}$\\
\hline
\end{tabular}
\begin{tabular}{p{15cm}}
Note: For KOI-13, the results are from a model fit including a 3$\phi$ term.
\end{tabular}
\label{tab:alb}
\end{table*}
\indent If the phase function is composed solely of reflected light the planet's albedo can be described by
\begin{eqnarray}
F_{\mathrm{ecl}} = A_{\mathrm{g}} \left( \frac{R_{\mathrm{p}}}{a} \right)^2
\end{eqnarray}
where $A_{\mathrm{g}}$ is the geometric albedo. Based on the eclipse depth and assuming that there is no contribution from thermal emission, we calculate an albedo of less than 1 for all planets, except KOI-2133 (see Tables~\ref{tab:res1}-\ref{tab:res2}). We consider this strong evidence for KOI-2133 being a self-luminous object and most likely not a planet. We note that the albedo calculated in this way should be considered as an upper-limit, since for all these objects thermal emission can contribute significantly (see below). \\
\indent Previous observations of hot Jupiters indicate low albedos at optical wavelengths (e.g.~\citeauthor{Collier2002}~\citeyear{Collier2002};~\citeauthor{Leigh2003}~\citeyear{Leigh2003};~\citeauthor{Rowe2006}~\citeyear{Rowe2006};~\citeauthor{Cowan2011}~\citeyear{Cowan2011} for an ensemble of planets), consistent with theoretical models (\citeauthor{Burrows2008}~\citeyear{Burrows2008}). \\
\indent The albedo plays a direct role in the planet's equilibrium temperature, $T_{\mathrm{eq}}$, which can be calculated using the method of~\citet{LopezMorales} as
\begin{eqnarray}\label{eq:teq}
T_{\mathrm{eq}} = T_{\star} \left( \frac{a}{R_{\star}} \right)^2 [f(1-A_{\mathrm{B}})]^{1/4} 
\end{eqnarray}
where $A_{\mathrm{B}}$ is the Bond albedo, which, if we assume Lambert's law, we can be defined as $A_{\mathrm{B}}=\frac{3}{2}A_{\mathrm{g}}$. The re-radiation factor, $f$, has two extremes, $f$=1/4, corresponding to homogeneous re-distribution of energy across the planet, and, $f$=2/3, for instant re-radiation from the day-side, resulting in a very hot day-side and cold night-side. Although these two limiting cases are useful when calculating the equilibrium temperature, the true $f$ lies somewhere in between. The equilibrium temperature can be compared to the brightness temperature $T_\mathrm{B}$, the temperature of a black-body with the equivalent flux in the band-pass, which can be calculated as
\begin{eqnarray}
F_{\mathrm{ecl}} &=& \left( \frac{R_{\mathrm{p}}}{R_{\star}} \right)^2 \frac{\int B_{\lambda}(T_{\mathrm{B}}) T_{\mathrm{K}} \mathrm{d}\lambda}{\int (T_{\mathrm{K}} F_{\lambda,\star} \mathrm{d}\lambda)}
\end{eqnarray}
where $B_{\lambda}$ is the Planck function as a function of $T_{\mathrm{B}}$ and $T_{\mathrm{K}}$ and $F_{\lambda,\star}$ are as described in Sec.~\ref{sec:dopp}. This provides us with the brightness temperature of the planet's day-side. In addition, if we change $F_{\mathrm{ecl}}$ with $F_{\mathrm{n}}$, the flux from the planet's night-side, we can calculate the night-side brightness temperature $T_{\mathrm{B, night}}$. \\
\indent In the case of isothermal atmospheric emission, we would expect that $T_{\mathrm{B}}$ fall somewhere between $T_{\mathrm{eq,hom}}$ and $T_{\mathrm{eq,max}}$ and that $T_{\mathrm{B,night}}$ be less than $T_{\mathrm{eq,hom}}$. However, we find that for all planets, except TrES-2, the brightness temperature is actually greater than maximum equilibrium temperature and that, for TrES-2, KOI-64 and KOI-2133, the night-side temperature is greater than the homogeneous equilibrium temperature (see Tables~\ref{tab:res1}-\ref{tab:res2}). \\
\indent For KOI-2133, this, along with having an albedo $>$1, implies that is almost certainly a self-luminous object. For KOI-64, the very large discrepancy between the night-side and equilibrium temperature also suggests that it is most-likely self-luminous and not a planet. For TrES-2, the 1.2$\sigma$ difference is not significant, and can easily arise if the layers probed at optical wavelengths are at a higher temperature than the equilibrium temperature. \\
\indent Since KOI-13 and HAT-P-7 have a significant night-side flux detection, consistent with their homogeneous temperature, we can place a constraint on their maximum allowed albedo. This is calculated by assuming a uniform temperature across the planet's surface ($f$=1/4) equal to the night-side temperature derived from $F_{\mathrm{n}}$. For KOI-13 and HAT-P-7, we find a maximum albedo of 0.26 and 0.148, respectively. \\
\indent In general, the eclipse depths at optical wavelengths are likely a combination of reflected light and thermal emission. To investigate this we self-consistently solve for the eclipse depth as a function of $A_{\mathrm{g}}$ using
\begin{eqnarray}
F_{\mathrm{ecl}} = \left( \frac{R_{\mathrm{p}}}{R_{\star}} \right)^2 \frac{\int B_{\lambda}(T_{\mathrm{B,day}}) T_{\mathrm{K}} \mathrm{d}\lambda}{\int (T_{\mathrm{K}} F_{\lambda,\star} \mathrm{d}\lambda)} + A_{\mathrm{g}} \left( \frac{R_{\mathrm{p}}}{a} \right)^2 \nonumber \\
\end{eqnarray}
where we assume that $T_{\mathrm{B,day}}$=$T_{\mathrm{eq}}(A_{\mathrm{B}}=\frac{3}{2}A_{\mathrm{g}})$ as given in Eq.16. In the limit of $f$=1/4 (uniform temperature) this will provide an upper limit on $A_{\mathrm{g}}$ and a lower limit on $T_{\mathrm{B,day}}$. While if $f$=2/3, we will obtain the opposite. We find that for all planets, except KOI-2133, there is a physical solution that satisfies these equations (see Table~\ref{sec:alb}) and that all, except KOI-64, have albedos less than 0.3. \\
\indent For KOI-13, if we assume a homogeneous heat distribution, an albedo of, at most, 0.148 is needed to produce the observed night-side flux. Using this albedo limit, we calculate an expected day-side flux significantly lower than the observed day-side flux. However, this would not be a problem in the case where the emitting layers probed in the {\it Kepler} bandpass, are hotter than the equilibrium temperature, as inferred for CoRoT-2~\citep{Snellen2010}. For TrES-2, this is most-likely also the case. \\
\section{Conclusions}\label{sec:concl}
\indent We have presented new phase curves for five {\it Kepler} objects of interest (Kepler-5, Kepler-6, Kepler-8, KOI-64 and KOI-2133) and re-examined the phase curves of TrES-2, HAT-P-7 and KOI-13 using 15 quarters of {\it Kepler} data. \\
\indent The fitted and derived parameters, for each of these systems, can be found in Tables~\ref{tab:res1}-\ref{tab:res2}. The derived ellipsoidal masses of Kepler-5, Kepler-6, Kepler-8, TrES-2 and HAT-P-7 are within 3$\sigma$, of their published radial velocity measurements, while the derived Doppler mass for TrES-2 and HAT-P-7 is within 1$\sigma$ and 9$\sigma$, respectively. When we compared the ellipsoidal and Doppler amplitudes of HAT-P-7 and KOI-13 to five previous studies that listed uncertainty values, we found that our results were within 3$\sigma$, while our values for TrES-2 agreed with its two previous phase curve studies (See Sec.~\ref{sec:mass}).\\
\indent Our secondary eclipse and phase function values of Kepler-5, Kepler-8, KOI-64, TrES-2 agree with previous studies, while four of the six previous studies of HAT-P-7 are within 4$\sigma$ of our values. In addition, our eclipse depth for KOI-13 is within 4$\sigma$, to three previous studies, but our phase function amplitude differs greatly, partly due to contamination from KOI-13's companion. A previous study of Kepler-6 found an eclipse depth more than double our value, however a different number of observations and systematic removal method was used (See Sec.~\ref{sec:alb}). \\
\indent For KOI-13, in addition to the phase curve components described in Sec.~\ref{sec:analysis}, we measure an out-of-phase third cosine harmonic with an amplitude of $6.7\pm0.3$ ppm. We believe that this signal could be a perturbation of KOI-13's ellipsoidal variations caused by it's spin-orbit misalignment. \\
\indent For KOI-64 and KOI-2133, we derived planet masses, from ellipsoidal variations and Doppler boosting, of less than 6 $\mathrm{M_J}$. However, we found that their day- and night-side temperatures were much higher than their equilibrium temperatures and therefore they must be self-luminous objects. We conclude that KOI-64 and KOI-2133 are false-positives created by an eclipsing binary diluted by a third stellar companion or a fore- or background star within the same {\it Kepler} pixel. \\
\indent For the rest of the objects, we find albedos of less than 0.3, but conclude that for TrES-2 and KOI-13 it is likely that the atmospheric layers probed in the Kepler bandpass, are hotter than the equilibrium temperature, as inferred for CoRoT-2~\citep{Snellen2010}. \\
\acknowledgments
\indent We thank Marten van Kerkwijk for insightful discussions. This work was supported by grants to R.J. from the Natural Sciences and Engineering Research Council of Canada. E.d.M. is also supported in part by an Ontario Postdoctoral Fellowship. \\
\appendix
\begin{figure}[h]
\begin{center}
\scalebox{0.84}{\includegraphics{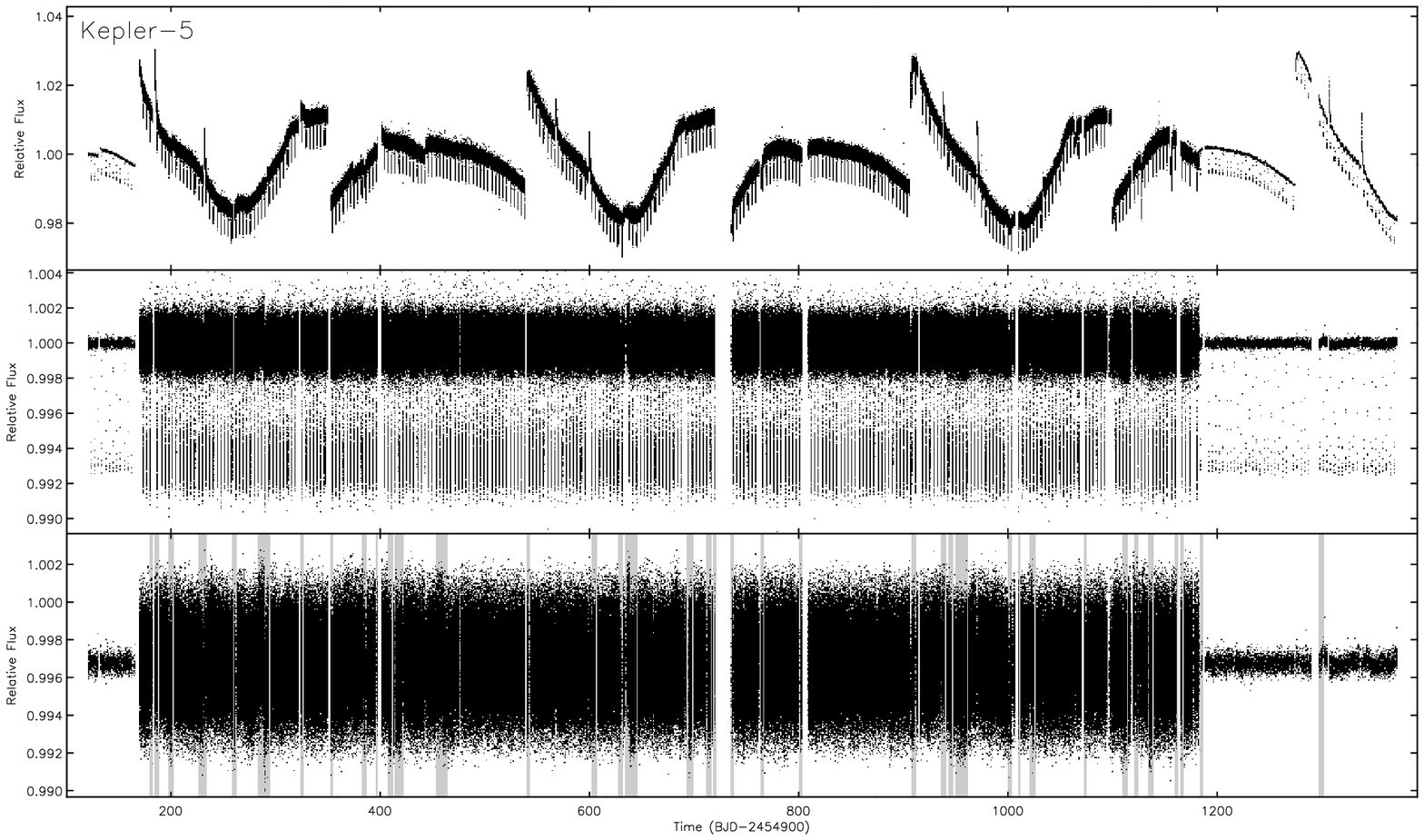}}
\scalebox{0.84}{\includegraphics{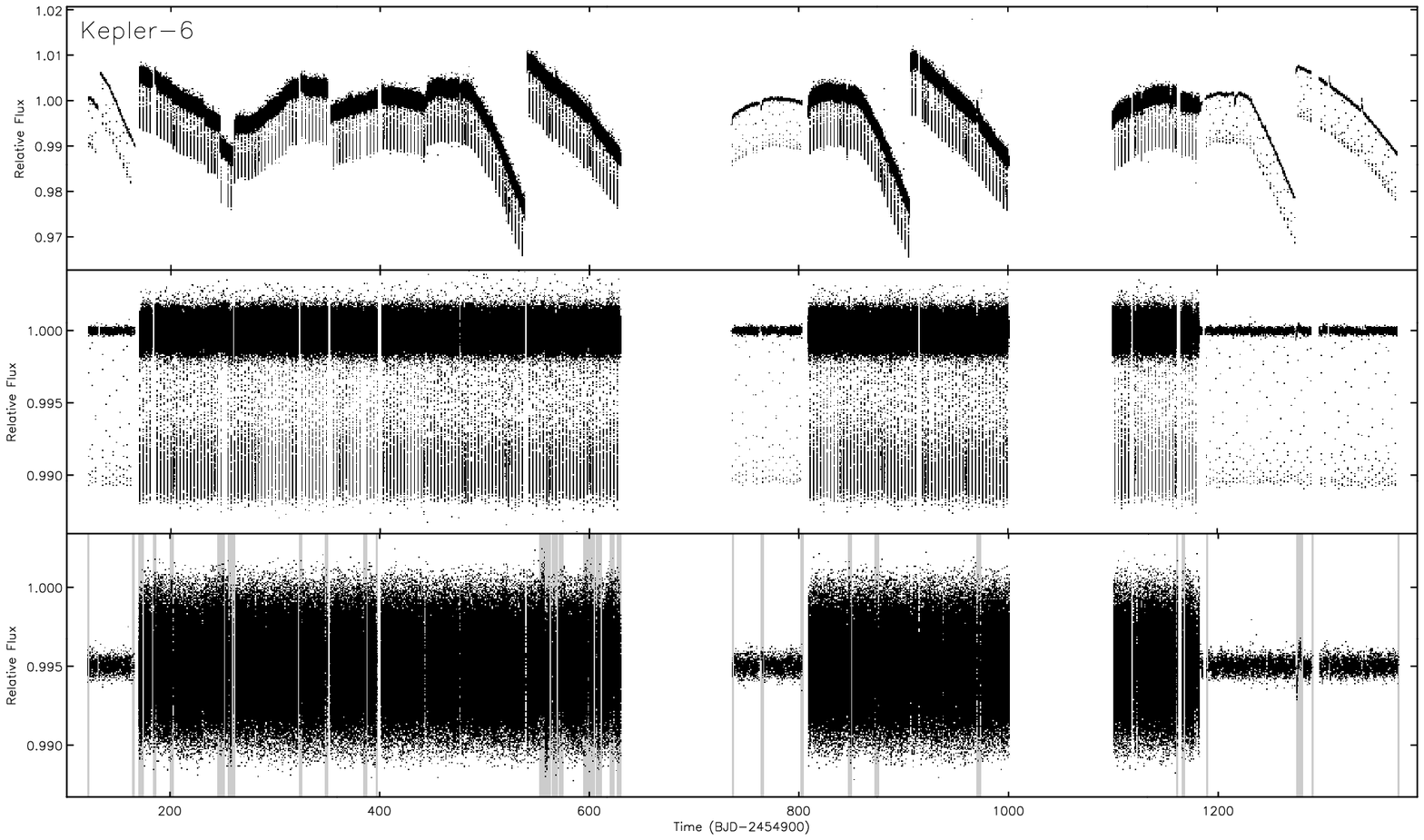}}
\end{center}
\caption{For Kepler-5 ({\it top} plot) and Kepler-6 ({\it bottom} plot), the {\it top} panel contains the raw SAP light curve, the {\it middle} panel is after cotrending and the {\it bottom} panel is after cotrending and removing the transits and outliers. The shaded portions indicate where we removed orbits because of a poor CBV fit.}
\label{fig:l1}
\end{figure}
\begin{figure}[h]
\begin{center}
\scalebox{0.84}{\includegraphics{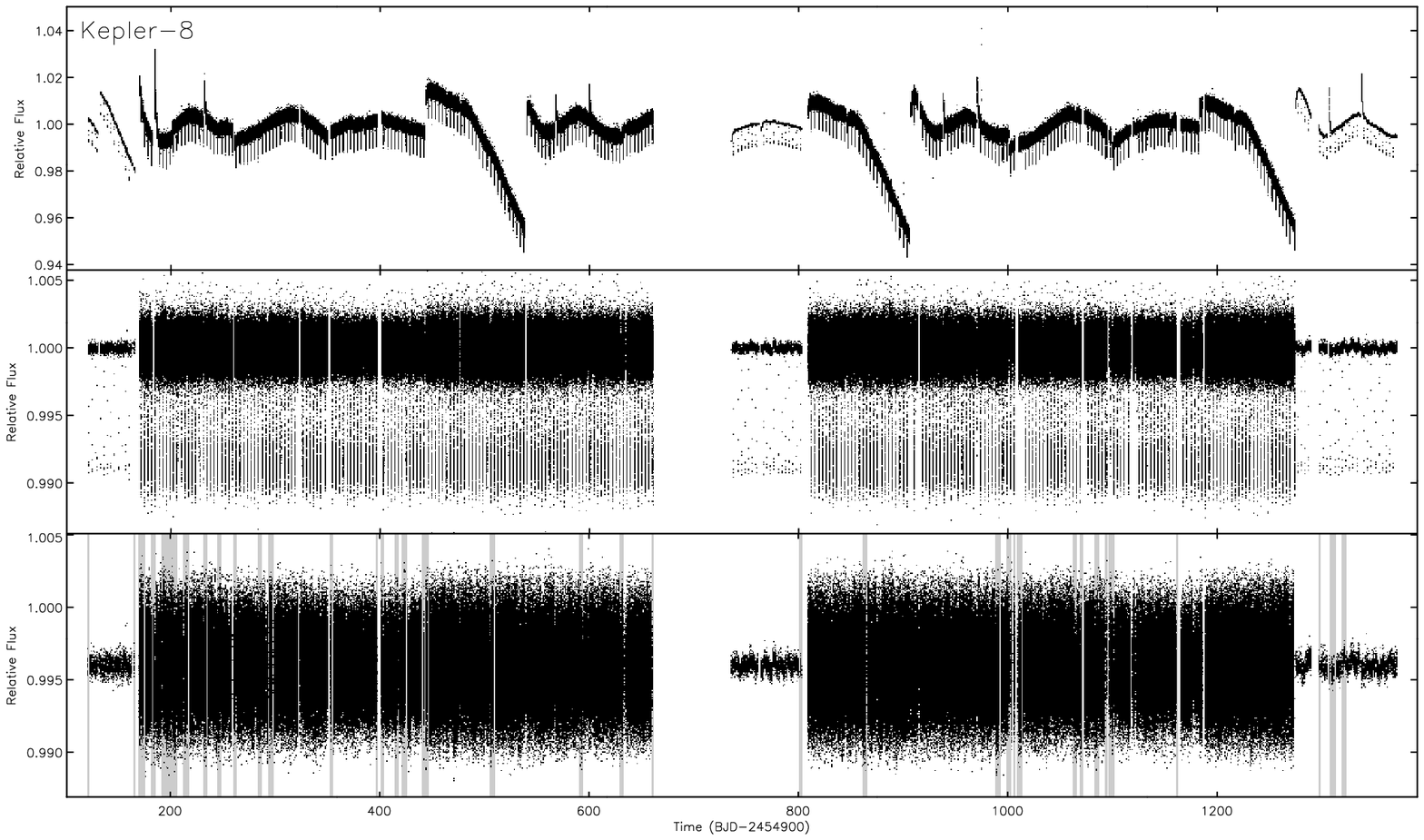}}
\scalebox{0.84}{\includegraphics{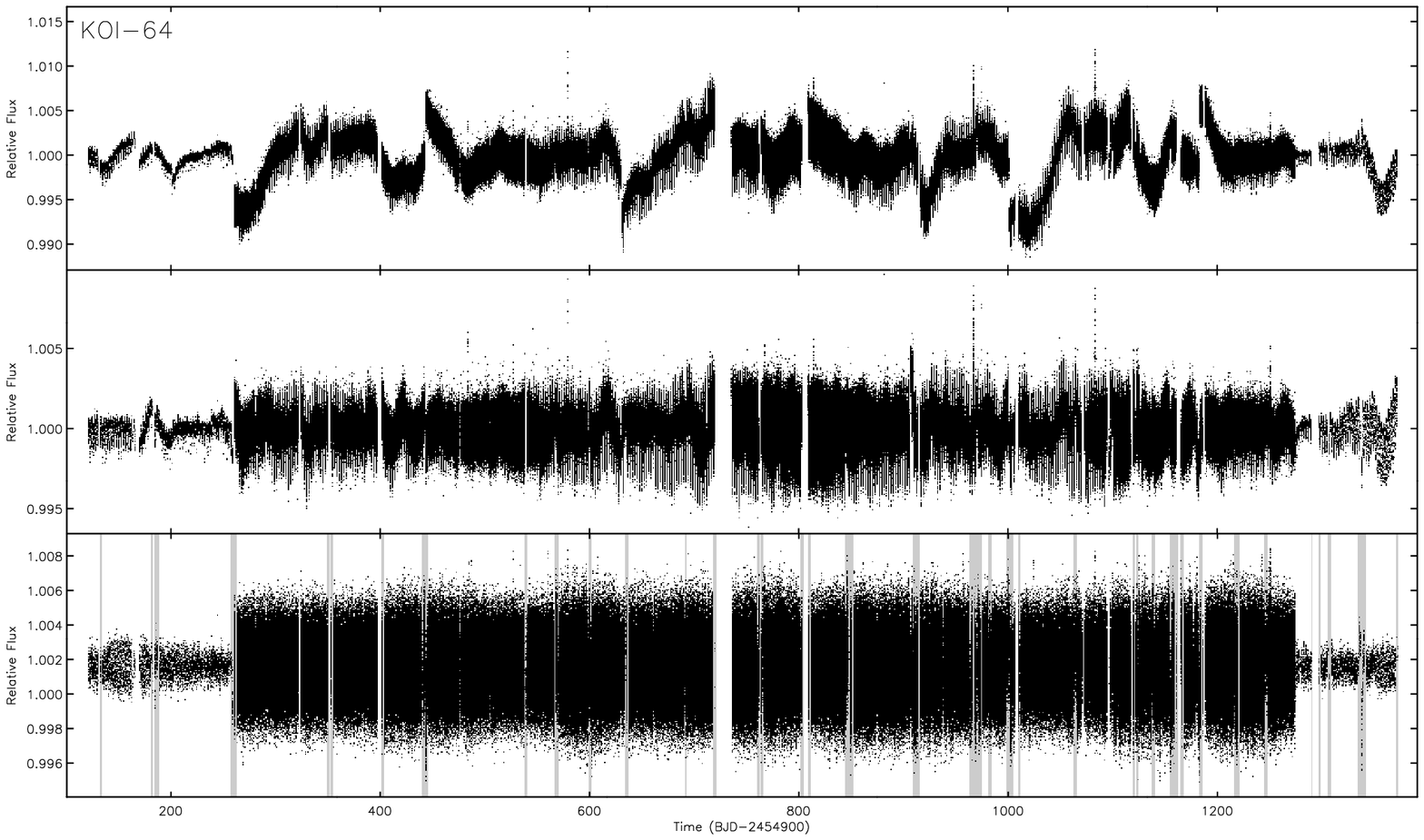}}
\end{center}
\caption{Same as Fig.\ref{fig:l1}, but for Kepler-5 ({\it top} plot) and KOI-64 ({\it bottom} plot) and where, for KOI-64, the {\it bottom} panel contains the cotrended/out-of-transit/outlier-filtered light curve after stellar variability removal (as described in Sec.~\ref{sec:var}).}
\label{fig:l2}
\end{figure}
\begin{figure}[h]
\begin{center}
\scalebox{0.84}{\includegraphics{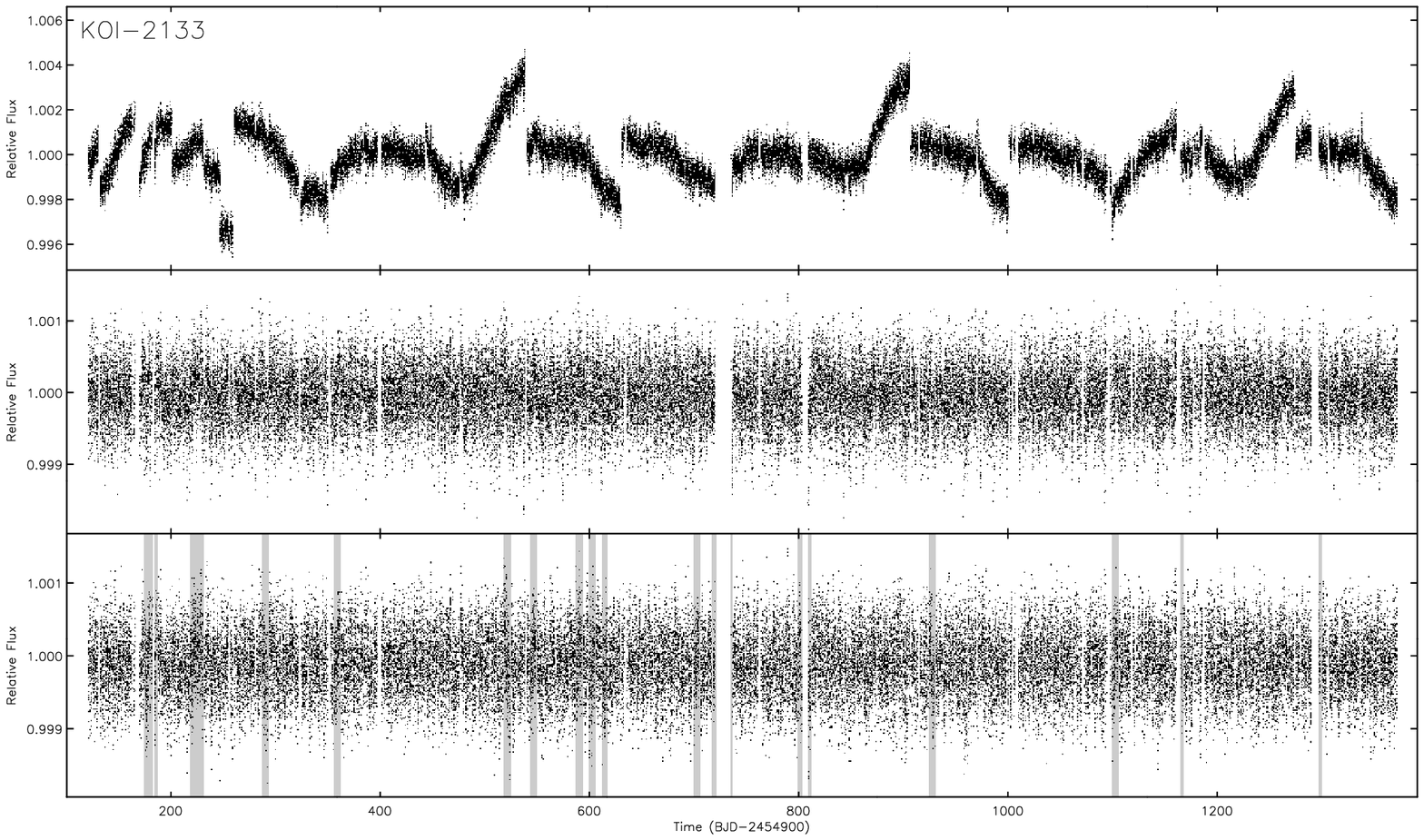}}
\scalebox{0.84}{\includegraphics{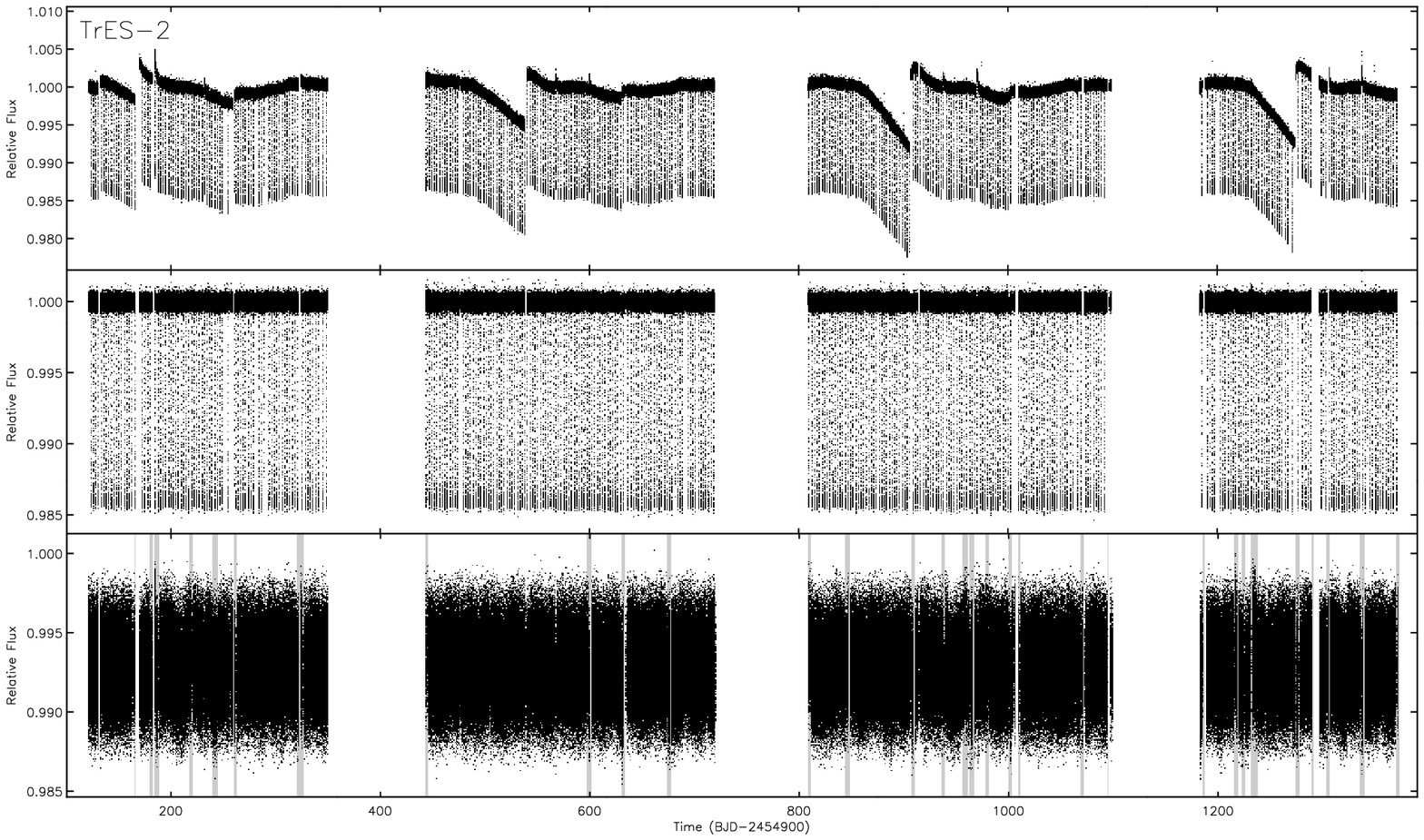}}
\end{center}
\caption{Same as Fig.\ref{fig:l1}, but for KOI-2133 ({\it top} plot) and TrES-2 ({\it bottom} plot).}
\label{fig:l3}
\end{figure}
\begin{figure}[h]
\begin{center}
\scalebox{0.84}{\includegraphics{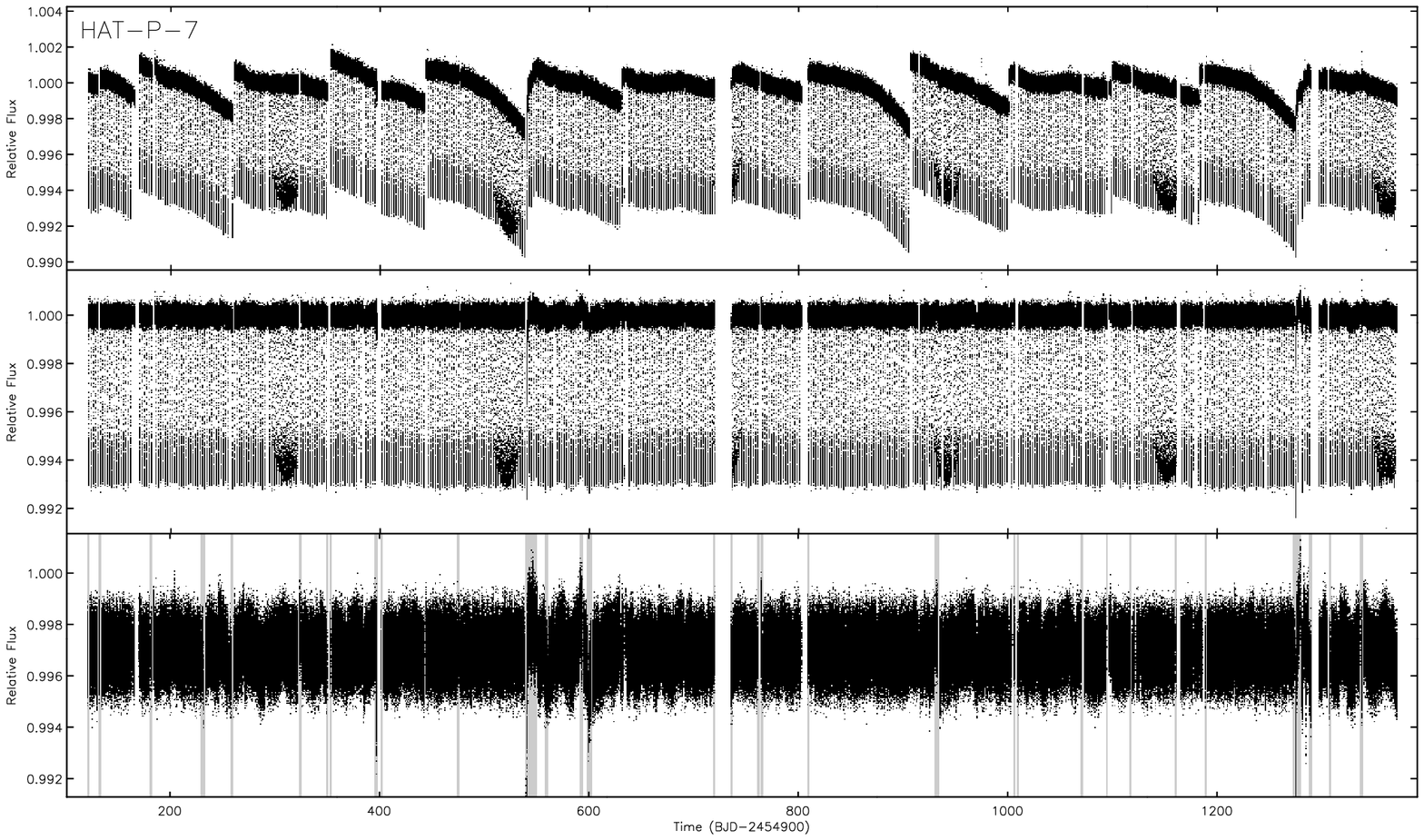}}
\scalebox{0.84}{\includegraphics{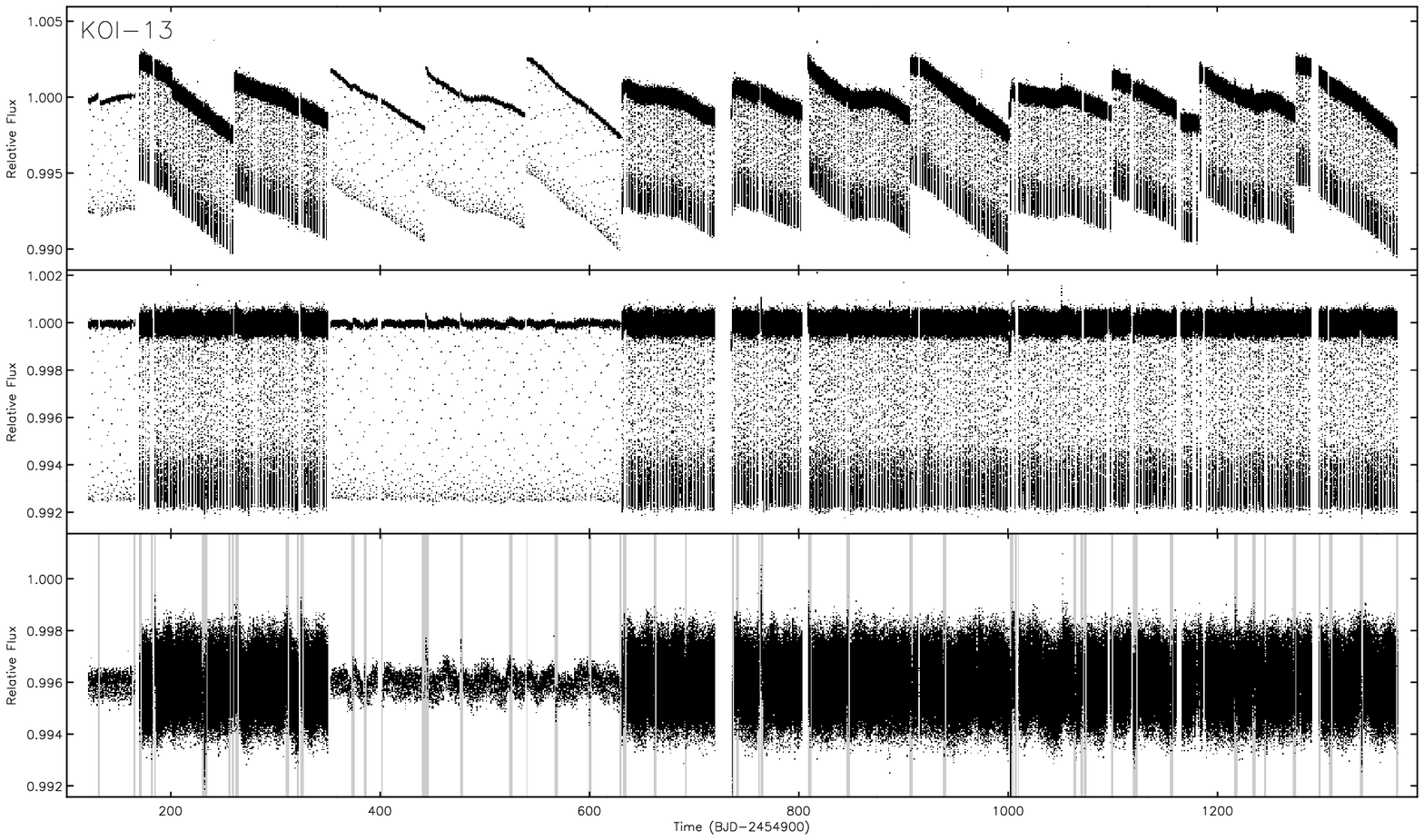}}
\end{center}
\caption{Same as Fig.\ref{fig:l1}, but for HAT-P-7 ({\it top} plot) and KOI-13 ({\it bottom} plot).}
\label{fig:l4}
\end{figure}
\end{document}